# Revisiting Homomorphic Wavelet Estimation and Phase Unwrapping

Roberto H. Herrera* and Mirko van der Baan

Department of Physics, University of Alberta, Edmonton T6G 2G7, CA.

rhherrer@ualberta.ca, Mirko.vanderBaan@ualberta.ca

**Summary**

Surface-consistent deconvolution is a standard processing technique in land data to uniformize the wavelet across all sources and receivers. The required wavelet estimation step is generally done in the homomorphic domain since this is a convenient way to separate the phase and the amplitude spectrum in a linear fashion. Unfortunately all surface-consistent deconvolutions make a minimum-phase assumption which is likely to be sub-optimal. Recent developments in statistical wavelet estimation demonstrate that nonminimum wavelets can be estimated directly from seismic data, thus offering promise to create a nonminimum phase surface-consistent deconvolution approach. Unfortunately the major impediment turns out to be phase unwrapping. In this paper we review several existing phase unwrapping techniques and discuss their advantages and inconveniences.

**Introduction**

We aim to create a new nonminimum-phase surface-consistent approach via log-spectral averaging by continuing research started by Tria et al. (2007). Nevertheless we have to deal previously with phase unwrapping problem as a vital step in log-spectral averaging and a long-standing problem in homomorphic deconvolution (Oppenheim et al., 1976; Lines, 1976; Tribolet, 1977). In this paper first we give a brief introduction to homomorphic deconvolution. We then provide the results of the performance of four methods when deal with common seismic situations, time shifting and phase rotation. Finally we compare these phase unwrapping techniques and discuss their advantages and inconveniences.

**Theory and Method**

The seismic trace is generally given as a convolution between the propagating wavelet and the reflectivity series of the earth and normally it is assumed that a white noise is added to the trace (Angeleri, 1983).

$$s(t) = w(t) \ast r(t) + \eta(t), \qquad (1)$$

where the observation $s(t)$ is the output of a linear system composed by the reflectivity function $r(t)$ and an unknown impulse response, denoted as the source wavelet $w(t)$, so we are addressing a blind deconvolution problem. The convolutional operator is denoted by $\ast$ and $\eta(t)$ is the additive noise.

In the frequency domain equation (1) can be expressed as:

$$S(f) = W(f)R(f) + N(f), \qquad (2)$$

where $S(f)$, $W(f)$ and $N(f)$ are the Fourier transforms of $s(t)$, $w(t)$ and $\eta(n)$ respectively.

A signal can be transformed from the time domain to the complex-valued cepstral domain by Fourier transformation ($FT\{\}$), followed by a log operation:

$$\hat{S}(f) = \log(S(f)) = \log(|S(f)| e^{j \arg[S(f)]}) = \log|S(f)| + j \arg[S(f)], \qquad (3)$$

with arg[$S(f)$] denoting the continuous phase of the complex value $S(f)$.

Then the cepstral transform of $s(t)$ is given by $\hat{s}(t) = \text{FT}^{-1}\{\hat{S}(f)\}$. Assuming that the noise content $N(f)$ in (2) is negligible the convolution model in (1) is transformed into additions in the cepstral domain:

$$\hat{s}(t) \approx \hat{w}(t) + \hat{r}(t) . \quad (4)$$

It is expected that for short band limited sources, $\hat{w}$ tends to concentrate near the origin while $\hat{r}$ is spread out over the higher quefrency values. For a single trace assuming a random white reflectivity sequence and a constant wavelet, the model (4) should work. But in practical cases more traces are needed to stabilize the performance in the presence of noise.

Spatial averaging in the log-spectral domain was proposed by Clayton and Wiggins (1976) and Otis and Smith (1977) for wavelet estimation and has been employed in surface-consistent deconvolution (SCD) by Taner and Koehler (1981). Averaging the log-spectrum is a suitable avenue for both methods namely wavelet estimation in the homomorphic domain and SCD. Unfortunately the complex logarithm is a multivalued function forcing the cepstral averaging technique to correct phase unwrapping first.

The phase spectrum of the observe signal $s(t)$ is often estimated as:

$$ARG[S(f)] = \arctan(\textbf{Im}\{S(f)\} / \textbf{Re}\{S(f)\}) , \quad (5)$$

where $ARG[S(f)]$ is the principal value of the phase obtained by the arctangent function of the Fourier transform of $s(t)$. This principal value is restricted to the range $[-\pi, \pi)$. The unwrapping process deals with obtaining samples of the unwrapped phase arg[$S(f)$] from samples of the phase modulo $2\pi$, $ARG[S(f)]$. One approach of phase unwrapping is based on the relation, $\arg[S(f)] = ARG[S(f)] + 2\pi n(f)$, with $n(f)$ denoting an integer that determines the appropriate multiple of $2\pi$ to add or subtract to the principal value $ARG[S(f)]$ ensuring the continuity of the unwrapped phase arg[$S(f)$].

Shatilo (1992) compared six methods with application to seismic phase unwrapping finding they are all likely to fail in complex cases. We test three methods reported after this previous work and include the one based on integration of the phase derivative (Stoffa et al., 1974) for comparison.

Our phase unwrapping survey includes the Matlab unwrap function based on the $2\pi$ jump correction, i.e., removing discontinuities introduced by the arctangent function in (5), called in this work PU-M method. The second method considered is the Stoffa's method (Stoffa et al., 1974) which aims to ensure continuity of the unwrapped phase, by numerical integration of the derivative of the unwrapped phase.

Steiglitz and Dickinson (1977), suggested to compute the unwrapped phase by adding together the phase contribution of each root of the Z-transform. Thus we need a polynomial factorization (PF) of the Z-transform of $s(t)$ to perform the phase unwrapping without the use of the arctangent function. We will take advantage of the PF algorithm proposed by Treitel et al. (2006) to find the roots and then perform the addition of the phase of each root. This method is referenced as PU-F.

The last method is called PU-K, Kaplan's method (Kaplan and Ulrych, 2007), where the wrapped phase is transformed to the complex plane (named w-plane). Then instead of looking for jumps in the actual phase this method seeks for zero crossings in the real negative axis in the complex domain. The algorithm operates in this domain, since the $2\pi$ jumps are naturally identified by a branch-cut allowing for easy identification.

**Examples**

We compared the performance of four phase unwrapping algorithm (PU-M: Matlab , PU-K: Kaplan's Method, PU-F: Polinomial Factoring, and PU-S: Stoffa's Method based on two common sources of phase wrapping (1) time shifting and (2) constant-phase rotation. A 90º rotated Ricker wavelet was shifted by 30, 60 and 90 ms, as is shown in Figure 1 top. The time shift implies a phase change, i.e. more jumps in the wrapped phase (Figure 1 middle). Finally in Figure 1 bottom, the ideal phase unwrapped shows the correspondence between the slope and the time displacement.

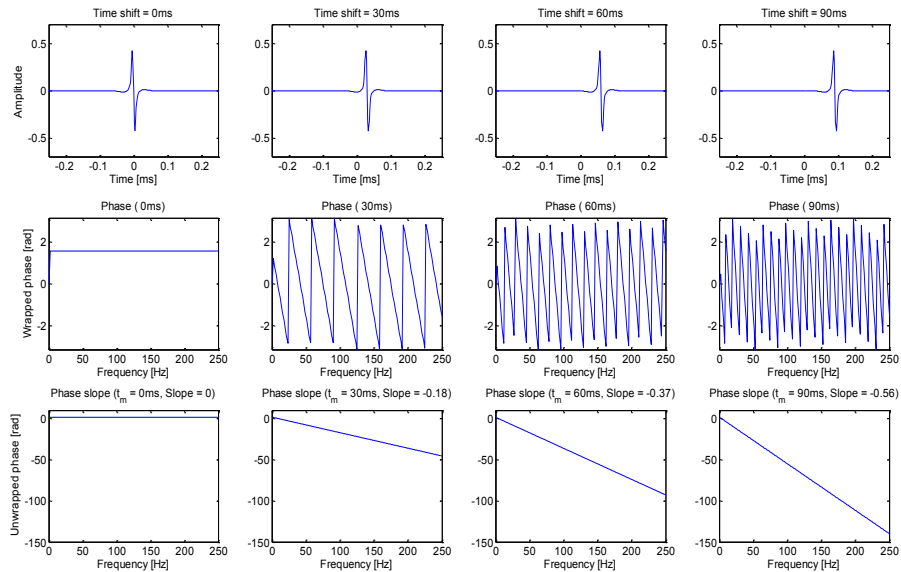

Figure 1: 90°-rotated Ricker wavelet and time shifted by 30, 60 and 90 ms (top). The correspondent wrapped phase (middle). The unwrapped phase showing the slope (bottom).

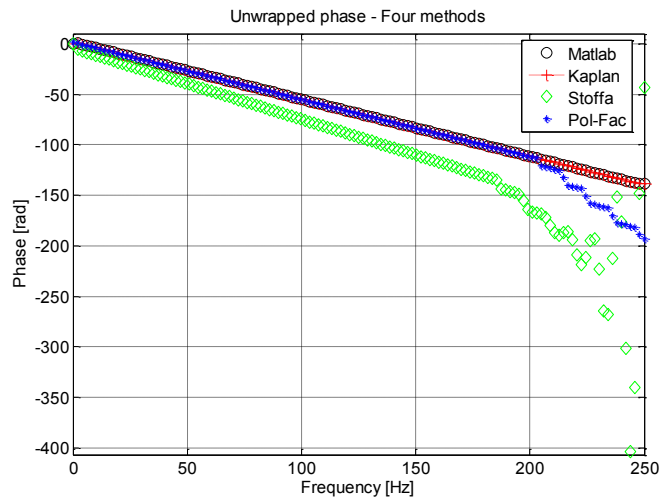

Figure 2: Phase unwrapping of a 90°-rotated Ricker wavelet with 90 ms time shift. The diamond curve represents the unreliable result of Stoffa's method. The other methods are able to estimate both the constant phase rotation and the time shifting.

Figure 2 shows the unwrapped phase by this four methods for the 90°-rotated Ricker wavelet with 90 ms of time shift. The Stoffa's method based on integration of the phase derivative shows unreliable results, while the rest produces accurate estimates of both the constant phase rotation and the linear phase due to the time shifting, except near the Nyquist frequency due to lack of signal.

Finally we tested the performance of these algorithms by changing the sample rate and reducing the number of samples for fixed total time length (see Table 1).

The Stoffa's method fails to give the exact value in all cases. This is in accordance with what was found by Shatilo (1992). The appropriated tuning of the spectral zone (Angeleri, 1983) could solve the problem but this will impose an additional task to the unwrapping scheme in every case.

Table 1: Phase estimation of a 90º-rotated Ricker wavelet with 90 ms time shift.

| No-samples | G-T | PU-M | PU-K | PU-F | PU-S |
|---|---|---|---|---|---|
| 256 | 90 | 88.6 | 89.3 | 90.0 | -235.0 |
| 512 | 90 | 89.3 | 89.3 | 90.0 | -31.0 |
| 1024 | 90 | 90.3 | 90.3 | 90.0 | 78.0 |
| Cpu-time(1024)[s] | - | 0.016 | 0.016 | 1.02 | 0.86 |

The Polynomial factorization (PU-F) was exact in phase estimation, reproducing the true phase in all cases, where the number of samples adequately represents the wavelet. Nevertheless it is computationally expensive and may be prohibitively expensive for large data sets.

There is a similar performance between Kaplan's and Matlab's methods with a slightly difference for the case of 256 samples. In the last row of Table 1, we show that PU-F is the most time consuming method, calculated with a signal of 1024 samples.

Matlab's and Kaplan's methods produce a good trade-off between performance and computational cost. We prefer the mathematical foundation of Kaplan's method and therefore recommend this approach.

## Conclusions

Robust phase unwrapping is a fundamental step in homomorphic wavelet estimation and deconvolution. Methods based on polynomial factorization are accurate but expensive. Approaches based on looking for discontinuous jumps larger than $\pi$ in wrapped phases work well as does Kaplan's method of searching for real-valued zero-crossings in the complex w-plane. We recommend either of the latter two techniques.

## Acknowledgements

The authors thank the BLISS sponsors for financial support. We are, also gratefully to S. Kaplan for kindly supplying his Matlab code for the w-plane method.